\documentclass[prb,twocolumn,showpacs]{revtex4}
\usepackage{graphicx}
\usepackage{amsmath}
\usepackage{amssymb}
\usepackage{dcolumn}
\usepackage{float}
\usepackage{bm}

\DeclareMathAlphabet{\bi}{OML}{cmm}{b}{it}

\def\be{\begin{equation}}
\def\ee{\end{equation}}
\def\bearr{\begin{eqnarray}}
\def\eearr{\end{eqnarray}}
\def\la{\langle}
\def\ra{\rangle}

\begin{document}
\title{Zitterbewegung of electrons 
in quantum wells and dots in presence of an in-plane magnetic field }
\bigskip
\author{Tutul Biswas and Tarun Kanti Ghosh\\
\normalsize
Department of Physics, Indian Institute of Technology-Kanpur,
Kanpur-208 016, India}
\date{\today}
 
\begin{abstract}
We study the effect of an in-plane magnetic field on the 
$ zitterbewegung$ (ZB) of electrons in a semiconductor quantum 
well (QW) and in a quantum dot (QD) with the Rashba and Dresselhaus 
spin-orbit interactions.
We obtain a general expression of the time-evolution of the position
vector and current of the electron in a semiconductor quantum well. 
The amplitude of the oscillatory motion is directly related to the
Berry connection in momentum space. 
We find that in presence of the magnetic field the ZB in 
a quantum well does not vanish when the strengths of the Rashba 
and Dresselhaus spin-orbit interactions are equal.
The in-plane magnetic field helps to sustain the ZB in quantum 
wells even at low value of $k_0 d$ (where $d$ is the width of the
Gaussian wavepacket and $k_0$ is the initial wave vector).
The trembling motion of an electron in a semiconductor quantum well 
with high Lande g-factor (e.g. InSb) sustains over a long time,
even at low value of $k_0 d$. 
Further, we study the ZB of an electron in quantum dots within the two sub-band 
model numerically. The trembling motion persists in time even when the
magnetic field is absent as well as when the strengths of the SOI are equal.
The ZB in quantum dots is due to the superposition of oscillatory 
motions corresponding to all possible differences of the energy eigenvalues
of the system. This is an another example of multi-frequency ZB phenomenon.

\end{abstract}

\maketitle

\section{Introduction}
In recent years there is a growing interest in the field of spin based
electronic devices. There has been a lot of study in this field after 
the proposal of the spin field effect transistor by 
Datta and Das \cite{datta}. The charge carriers carry spin in 
addition to their charges. 
The ultimate goal of this field is to control the spin degree of 
freedom of the charge carriers to produce and detect
spin-polarized current in semiconductor nanostructures \cite{fabian}.
One can develop device technology \cite{wolf} and quantum information 
processing \cite{david} in future on the basis of the manipulation 
of the spin degree of freedom. 
The coupling between intrinsic spin of an electron with its orbital 
angular momentum constitutes the intrinsic spin-orbit interaction (SOI) 
in low-dimensional semiconducting systems. Particularly, there are two kind 
of SOI present in low-dimensional semiconductor structures. 
One is the Rashba \cite{rashba} spin-orbit interaction (RSOI)
and another is the Dresselhaus \cite{dress} spin-orbit interaction (DSOI). 
The RSOI arises mainly from the inversion asymmetry of the confining potential 
in semiconductor heterojunctions. The strength of RSOI is proportional to the 
electric field externally applied which can be tuned by an external bias 
\cite{nitta,mats} or internally generated due to the crystal potentials.
On the other hand DSOI is present in bulk materials and semiconductor 
heterostructures which lack bulk inversion symmetry. The form of DSOI term 
strongly depends on the growth direction of semiconductor quantum well (QW)
\cite{car,chen,andr}. 
The strength of DSOI depends on properties of the material and the crystal 
structure.

In 1930, Erwin Schrodinger \cite{Schro} predicted that a 
free particle described by the relativistic Dirac equation 
will perform an oscillatory motion which is known as the 
$zitterbewegung$ (ZB). 
The free relativistic Dirac particle
has two energy branches:
$\epsilon({\bf p}) = \pm \sqrt{{{{m}_0}^2}c^4+{c^2}{p^2}}$.
It is well understood that this oscillatory motion 
results from the interference between these two energy branches
\cite{huang}. 
The large oscillation frequency $\omega_{_c} \simeq 10^{21}$ Hz
and the small oscillation amplitude $ \lambda_c \simeq 10^{-13}$ m
is not accessible to the modern experimental techniques.  
Most of the studies on the ZB of electrons used
plane waves to describe the electrons.
It was pointed out by Lock \cite{lock} that
plane wave is not a localized state and therefore, rapid oscillations
on the average position of a plane wave has some limitations.
He also demonstrated that when an electron is described by a wave packet,
the ZB oscillation has a transient character.

In 2005, Zawadzki et al. \cite{zawadki} studied
ZB in narrow gap semiconductor (NGS) by using the analogy between
$\bf k \cdot \bf p$ theory of the energy bands in NGS and the free Dirac
relativistic equation for electrons. They found much more
favorable amplitude and frequency of the oscillation than those in 
a vacuum for a free electron. 
At the same time, Schliemann et al. \cite{john,loss} has studied the ZB of 
an electron in III-V zincblende semiconductor QWs in the 
presence of the SOI thoroughly.
The above studies initiated intense theoretical research on
the ZB of electrons in various condensed matter systems 
\cite{review,demi,rusin,rusin1,maksinova,winkler,lamata,nature}. 

The proposal of an experimental scheme for observing ZB in
ultra-cold \cite{clark} atomic gases is also given.
The trembling motion was proposed for photons in a two-dimensional
photonic crystal and for Ramsey interferometry \cite{photonic}.
It was reported in Ref. \cite{sonic} that an acoustic analog of 
ZB in a macroscopic two-dimensional sonic crystal was observed.
A general theory of ZB of a multi-band Hamiltonian is studied by 
David and Cserti \cite{cserti}.
Recently, Vaseghi \cite{vaseghi} et al.  
studied the effect of an external perpendicular magnetic 
field on the ZB for both
quantum wire and quantum dot within the two-sub-band model. 

The dimensionless parameter $ p_0 = k_0 d$ dictates whether the
motion will be oscillatory or not. It was shown in Ref. [14] that
the motion will be oscillatory when $p_0 \gg 1 $. It was also shown
that the ZB vanishes when the strengths of the RSOI and DSOI are equal. 
This is due to an additional conserved quantity.

In this work we study effect of the in-plane magnetic field on the
ZB of electrons in a semiconductor QW and in a QD with the Rashba and 
Dresselhaus SOI.
We obtain an analytical expression of the time-evolution of the position
vector of an electron in a QW by using the Schrodinger picture.
The advantage of using the Schordinger picture is to see the 
direct relation between the amplitude of the oscillatory motion 
and the Berry connection in momentum space.
We find that in presence of the magnetic field the ZB in
a quantum well does not vanish even when the strengths of the Rashba
and Dresselhaus spin-orbit interactions are equal.
The ZB in a QW sustains even at very low value of $k_0 d$ 
due to the presence of the magnetic field.
The trembling motion of an electron in a semiconductor QW
with high Lande g-factor (e.g. InSb) sustains over a long time,
even at very low value of $k_0 d$.
The time-period of the oscillation decreases as the magnetic field strength
is increased.
Next, we study the ZB of an electron in a QD within the two sub-band
model numerically. The trembling motion does not fade away in time even 
when the magnetic field is absent as well as when the strength of the SOIs 
is equal. 

This paper is organized as follows. In section II, we review the Hamiltonian
and eigenstates of a 2DEG with both type of SOI in the presence of an 
in-plane magnetic field.
In section III, we derive time-evolution of the electron' position
vector in the Schordinger picture.
The numerical results and discussions are given in section V.  
Section VI contains the calculation and 
result of ZB of electrons in a GaAs/AlGaAs QD. 
The conclusion is presented in Section VII.

\section{Two-dimensional electron gas with spin-orbit interactions}
We consider a 2DEG with the Rashba and Dresselhaus spin-orbit 
interactions in presence of an in-plane magnetic field 
${\bf{B}} = \hat{x}{B}_x + \hat{y} {B}_y$.
The single-particle Hamiltonian of this system is given by
\begin{eqnarray}
H &=& \frac{{\bf p}^2}{2m^\ast} + 
\frac{\alpha}{\hbar}\big({\sigma}_x{p}_y-{\sigma}_y{p}_x\big)
+\frac{\beta}{\hbar}\big({\sigma}_x{p}_x-{\sigma}_y{p}_y\big) \nonumber\\
&+&\frac{g^\ast}{2}{\mu_{_B}}\big(B_x\sigma_x+B_y\sigma_y\big),
\end{eqnarray}
where ${\bf p} $ and $m^\ast$ is the momentum and the effective mass 
of an electron, respectively. Here $ \alpha $ and $ \beta $ are the 
strengths of the Rashba and Dresselhaus spin-orbit interaction. 
The last term in the Hamiltonian is the Zeeman term due to the
application of the in-plane magnetic field $\bf B$, $g^\ast$ 
is the effective Lande-g factor and $\mu_{_B}$ is the Bohr magneton.
The energy eigenvalues and the corresponding eigenstates of the 
system \cite{chang,rod} are, respectively, given by
\begin{equation}
E_{\pm}({\bf k})=\frac{\hbar^2\big(k_x^2+k_y^2\big)}{2m^\ast}\pm \sqrt{C^2+D^2},
\end{equation}
where $ C= \alpha k_y + \beta k_x + g^\ast\mu_{_B} B_x/2 $, 
$D = \alpha k_x + \beta k_y - g^\ast \mu_{_B} B_y/2 $,
and the spin eigenstates are
\begin{eqnarray} \label{spinstates}
\vert k,+\ra=\frac{1}{\sqrt{2}}\begin{pmatrix} 1
\\ -i e^{-i\theta_k} \end{pmatrix}, \hspace{0.1in} 
\vert k,-\ra=\frac{1}{\sqrt{2}}\begin{pmatrix} 
-ie^{-i\theta_k}\\1\end{pmatrix},
\end{eqnarray}
with $ \theta_k = \tan^{-1}(C/D) $.

In absence of the magnetic field, a new conserved quantity
$ \Sigma = \sigma_x - \sigma_y$ exists for $\alpha = \beta$
\cite{john_prl,john_prb}.
In this situation, the spin eigenstates $\vert k, \pm \ra $ become 
independent of the wave vector $k$ and therefore the ZB does not 
exist due to the absence of spin randomization \cite{john,loss}.
The situation is completely different when the in-plane
magnetic field is present. It is clear from Eq. 
(\ref{spinstates}) 
that the spin states are always function of $k$ even at $\alpha = \beta$
as long as ${\bf B} \neq 0$. Therefore, spin randomization occurs and we would 
expect to see the ZB which is shown in the next section.

\subsection{Time-evolution of the wave packet and the Zitterbewegung}
We shall use the Schordinger picture to analyze the time-evolution of 
the electron position vector.
We represent the initial wave function of an electron by a Gaussian 
wave packet with initial spin polarization along the $z$ axis
as given by
\begin{eqnarray}
\psi\big({\bf r},0 \big)=\frac{1}{2\pi}\int\,d^2k a({\bf k}, 0)
e^{i\bf k \cdot r}
\left(\begin{array}{c}
1 \\0 \end{array}\right),
\end{eqnarray}
where
$a({\bf k},0) = d/(\sqrt{\pi}) e^{-\frac{1}{2} d^2\big({\bf k-k}_0 \big)^2}$.
Here, $ d$ and $ {\bf k}_0 $ is the initial width and the initial 
wave vector of the wave packet, respectively.
The time-evolution of the initial wave packet in the Schordinger
representation can be obtained in the usual manner as 
$\psi({\bf r}, t)= U(t) \psi({\bf r}, 0)$, where  
$ U(t) = e^{-iHt/\hbar} $ is the time-evolution operator.
After doing some straightforward algebra we obtain,
\begin{eqnarray} \label{time_wp}
\psi({\bf r}, t)&=&\frac{1}{2\pi}\int\ d^2k a({\bf k}, 0)e^{i\bf k \cdot r}
e^{-i\frac{\hbar k^2}{2m^\ast}t}\nonumber\\
&\times & \Big\{\cos[\omega({\bf k})t]\left(\begin{array}{c}
1 \\0 \end{array}\right)-e^{i\theta_k}\sin[\omega({\bf k})t]\left(\begin{array}{c}
0 \\1 \end{array}\right)\Big\},\nonumber\\
\end{eqnarray}
where
$\omega({\bf k})= \big[E_+({\bf k}) - E_-({\bf k})\big]/(2\hbar)$.

Equation (\ref{time_wp}) is nothing but the Fourier transformation of the 
following function:
$$ 
\phi({\bf k},t)=a({\bf k}, 0)e^{-i\frac{\hbar k^2}{2m^\ast}t}\left(\begin{array}{c}
\cos[\omega({\bf k})t]\\-e^{i\theta_k}\sin[\omega({\bf k})t]\end{array}\right).$$
The expectation value of position operator is then simply given by
\begin{eqnarray}
\la {\bf r}(t)\ra=i\int d^2k \phi^{\dagger}({\bf k},t) {\bf \nabla}_{\bf k}
\phi({\bf k},t).
\end{eqnarray}
Now it is straightforward to show that
\begin{eqnarray} \label{zb}
\la {\bf r}(t)\ra= \la {\bf r}(0)\ra&+&\frac{\hbar {\bf k}(0)}{m^\ast}t-\frac{1}{2}
\int d^2k \left|a({\bf k,0})\right|^2\Big({\bf \nabla}_{\bf k}\theta_k\Big)\nonumber\\
&\times&\Big\{1-\cos[\omega({\bf k})t]\Big\}.
\end{eqnarray}
The last oscillatory term $\cos[\omega({\bf k})t]$ in Eq. (\ref{zb}) 
indicates the ZB.
Two important observations can be made by analyzing Eq. (\ref{zb}).
First, the amplitude of the ZB is directly proportional to the 
Berry connection 
$\la k,\pm \vert i\frac{\partial}{\partial {\bf k}}\vert k,\pm \ra
= {\bf \nabla}_{\bf k}\theta_k $.
Second, for $\alpha = \beta $ the amplitude (or the Berry connection) of the
oscillation does not vanish when the in-plane magnetic field is present.
But it vanishes if the magnetic field is absent.

After substituting the expressions for 
$ a({\bf k},0)$, ${\bf \nabla}_{\bf k}\theta_k $ in Eq. (\ref{zb})
and considering the initial velocity of the wave packet is along 
$y$-direction (i.e. $k_{0x}=0, k_{0y} = k_0$)
one can easily obtain the following expressions

\begin{eqnarray} \label{zb1}
\la x(t) \ra & = &
\frac{d}{2\pi}e^{-d^2k_0^2}\int^{2\pi}_0d\phi
\int_0^\infty dq e^{-q^2+2dqk_{0y}\sin\phi} \nonumber \\ 
& \times &
\frac{q^2(1-\eta^2)\sin\phi+q(\varepsilon_x+\eta\varepsilon_y)}{Q^2}
\nonumber\\&\times&
\Big[1-\cos\Big(\frac{2\alpha}{\hbar d}tQ\Big)\Big],
\end{eqnarray}

and

\begin{eqnarray} \label{zb2}
\la y (t) \ra & = & \frac{\hbar k_{0y}}{m^\ast}t\nonumber\\
&+&\frac{d}{2\pi}e^{-d^2k_0^2}\int^{2\pi}_0d\phi
\int_0^\infty dq e^{-q^2+2dqk_{0y}\sin\phi} \nonumber \\ 
& \times &
\frac{q^2(\eta^2-1)\cos\phi+q(\eta\varepsilon_x+\varepsilon_y)}{Q^2}
\nonumber\\&\times&
\Big[1-\cos\Big(\frac{2\alpha}{\hbar d}tQ\Big)\Big],
\end{eqnarray}
where $ Q^2 = (q\cos\phi+\eta q\sin\phi-\varepsilon_y)^2 + (q\sin\phi
+\eta q\cos\phi+\varepsilon_x)^2 $,
$\eta=\beta/\alpha$, $q=kd$, 
$\varepsilon_x= g^\ast \mu_{_B} B_x d/(2\alpha)$ and 
$\varepsilon_y= g^\ast\mu_{_B} B_y d/(2\alpha)$.

From Eqs. (\ref{zb1}) and (\ref{zb2}) one can infer 
that ZB is absent in the limit $d \rightarrow 0$.
On the other hand, as $d \rightarrow \infty $ the Gaussian 
approaches to a delta function i.e. 
$a({\bf k},0)=\delta({\bf k}-{\bf k}_0)$ and in this limit we 
find an analytic expression for the ZB as given by
\begin{eqnarray}
\la x_{} (t)\ra & = &\frac{1}{2}\frac{(1-\eta^2)k_{0y}+
\frac{g^\ast\mu_{_B}}{2\alpha}(B_x+\eta B_y)}{(k_{0y}
+\frac{g^\ast\mu_{_B} B_x}{2\alpha})^2+(\eta k_{0y}-
\frac{g^\ast\mu_{_B} B_y}{2\alpha})^2}\nonumber\\
&\times&\Big\{1-\cos[\omega(k_{0y})t]\Big\},
\end{eqnarray}

and 

\begin{eqnarray} \label{zb4}
\la y_{} (t)\ra & = &\frac{\hbar k_{0y}}{m^\ast}t\nonumber\\
&+&\frac{1}{2}\frac{\frac{g^\ast\mu_{_B}}{2\alpha}(\eta B_x+ B_y)}{(k_{0y}
+\frac{g^\ast\mu_{_B} B_x}{2\alpha})^2+(\eta k_{0y}-
\frac{g^\ast\mu_{_B} B_y}{2\alpha})^2}\nonumber\\
&\times&\Big\{1-\cos[\omega(k_{0y})t]\Big\}.
\end{eqnarray}

 We also calculate the expectation values of the velocity of electron
in $x$ and $y$ direction. They are given by

\begin{eqnarray} \label{vel1}
\la v_x(t)\ra & = & \frac{\partial \la x \ra}{\partial t}\nonumber\\
& = & \frac{v_{_R}}{2\pi}e^{-d^2k_{0y}^2}\int^{2\pi}_0 d\phi 
\int_0^\infty dq e^{-q^2 + 2d q k_{0y} \sin\phi}\nonumber\\
& \times & \frac{q^2(1-\eta^2)\sin\phi+q(\varepsilon_x+\eta\varepsilon_y)}{Q}
\sin\Big(\frac{2\alpha Q}{\hbar d}t\Big), \nonumber \\
\end{eqnarray}

and

\begin{eqnarray} \label{vel2}
\la v_y(t)\ra & = &\frac{\hbar k_{0y}}{m^\ast}\nonumber\\
&+&\frac{v_{_R}}{2\pi}e^{-d^2k_{0y}^2}\int^{2\pi}_0 d\phi 
\int_0^\infty dq e^{-q^2 + 2d q k_{0y} \sin\phi}\nonumber\\
& \times & \frac{q^2(\eta^2-1)\cos\phi+q(\eta\varepsilon_x+\varepsilon_y)}{Q}
\sin\Big(\frac{2\alpha Q}{\hbar d}t\Big). \nonumber \\
\end{eqnarray}

Here, $v_{_R} = \Omega d=2\alpha/\hbar$ is the velocity corresponding to 
the RSOI. The corresponding currents are simply given by 
$\la j_x(t)\ra = e \la v_x(t)\ra$ and $\la j_y(t)\ra = e \la v_y(t)\ra$,
where $e$ is the electronic charge.
Eqs. (\ref{zb2}) and (\ref{vel2}) tell us that the ZB along $y$
direction vanishes when in-plane magnetic field and the DSOI are absent 
simultaneously \cite{john} because in this case
$\int_0^{2\pi} e^{2dqk_{0y}\cos{\phi}}\sin{\phi}d\phi=0$. But when there
is a finite in-plane magnetic field present the ZB in $y$ direction does 
not vanish even at $\beta=0$. 
So this is an effect of the in-plane magnetic field on ZB.

\subsection{Numerical Results and Discussion}
In this sub-section we evaluate time-evolution of the observables 
position vector and current density for different values of the
parameters like magnetic field, $k_0 d$, $\beta$ etc.\\

{\bf GaAs/AlGaAs QW}: 
We consider GaAs/AlGaAs quantum well for which the 
effective Lande $g$-factor is $g^\ast = -0.44 $. 
The value of the Rashba coefficient is 
taken to be $\alpha = 1.0 \times 10^{-11}$ eV-m.
We set the condition $dk_0=5$ in all the cases.  
To investigate the time dependence of the expectation values
of position and current of electron, we numerically evaluate
Eqs. ({\ref{zb1}}), ({\ref{zb2}}), ({\ref{vel1}}) and ({\ref{vel2}}).
Here it should be mentioned that Eqs. ({\ref{zb2}}) and
({\ref{zb4}}) contain two 
parts: the first depends linearly on time and the second one is 
oscillatory in time (responsible for ZB). The magnitude of the 
first term is very large compared to that of the second one. 
So if we plot this as a function of time we will get a straight line 
due to the dominating first term.

We consider three cases corresponding to three different values
of the parameter $\beta$ namely $\beta=0$, $\beta=0.5\alpha$ and
$\beta=\alpha$. For each case we plot $\la x(t)\ra/d$, $\la j_x(t)\ra/ev_R$
and $\la j_y(t)-j_0\ra/ev_R$ as a function of $\Omega t$ for different values
of magnetic field in Figs. [1-3]. Here we define the quantity $j_0$ as
$j_0=e\hbar k_{0y}/m^\ast$. From Fig. [1(a)], one can see that the amplitude of the ZB
increases as the magnetic field increases from its zero value. 
It is also noticeable that the ZB pattern is more oscillatory 
with increasing magnetic field. The current shows (in Figs. [1(b),1(c)]) the 
similar behavior as the position but it oscillates about zero. From Figs. [2] and [3]
we can see that as we increase the value of $\beta$, the amplitude of
ZB decreases. One important point is to be noted here that there is a 
definite phase difference between the currents in $x$ and $y$ direction 
when $\beta\neq\alpha$ as evident from Figs. [1] and [2]. 
But the situation is different when $\beta=\alpha$.
It can be seen from Fig. [3(b), 3(c)] the currents are oscillating in the 
same phase and this can be easily understood by analyzing Eqs. 
[\ref{vel1}, \ref{vel2}].

\begin{figure}[h]
\begin{center}\leavevmode
\includegraphics[width=105mm]{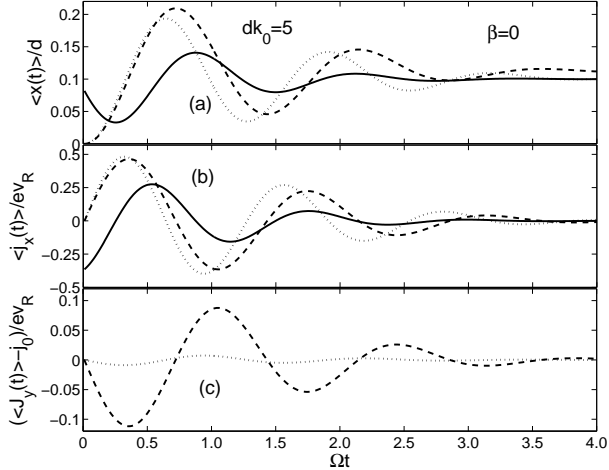}
\caption{Here $\la x(t)\ra/d$, $\la j_x(t)\ra/{ev_{_R}}$ and
$\la j_y(t)-j_0\ra/{ev_{_R}}$ are plotted as a function of
$\Omega t$. In the all the cases we set $\beta=0$ for GaAs/AlGaAs QW. 
Here, solid line: ($B_x=0$, $B_y=0$),
dotted line: ($B_x=1/\sqrt{2}$ T, $B_y=1/\sqrt{2}$ T),
dashed line: ($B_x=6$ T, $B_y=8$ T).}
\label{Fig1}
\end{center}
\end{figure}

\begin{figure}[t]
\begin{center}\leavevmode
\includegraphics[width=105mm]{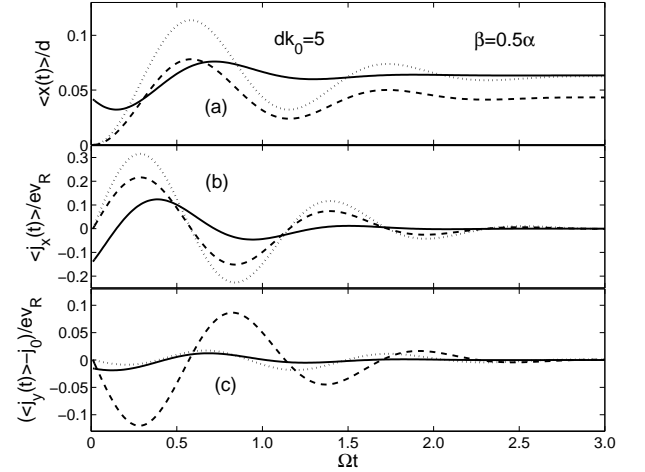}
\caption{Here $\la x(t)\ra/d$, $\la j_x(t)\ra/{ev_{_R}}$ and
$\la j_y(t)-j_0\ra/{ev_{_R}}$ are plotted as a function of
$\Omega t$. 
In all the cases we set $\beta=0.5\alpha$ for GaAs/AlGaAs QW. 
Here, solid line: ($B_x=0$, $B_y=0$),
dotted line: ($B_x=1/\sqrt{2}$ T, $B_y=1/\sqrt{2}$ T),
dashed line: ($B_x=6$ T, $B_y=8$T).}
\label{Fig2}
\end{center}
\end{figure}

In Ref. [14] it was shown that the ZB vanishes if $\alpha=\beta$ 
in the absence of any external magnetic field. But Fig. [3] shows that the ZB 
is still present when $\alpha=\beta$. This is due to the application of the
external in-plane magnetic field.

\begin{figure}[t]
\begin{center}\leavevmode
\includegraphics[width=105mm]{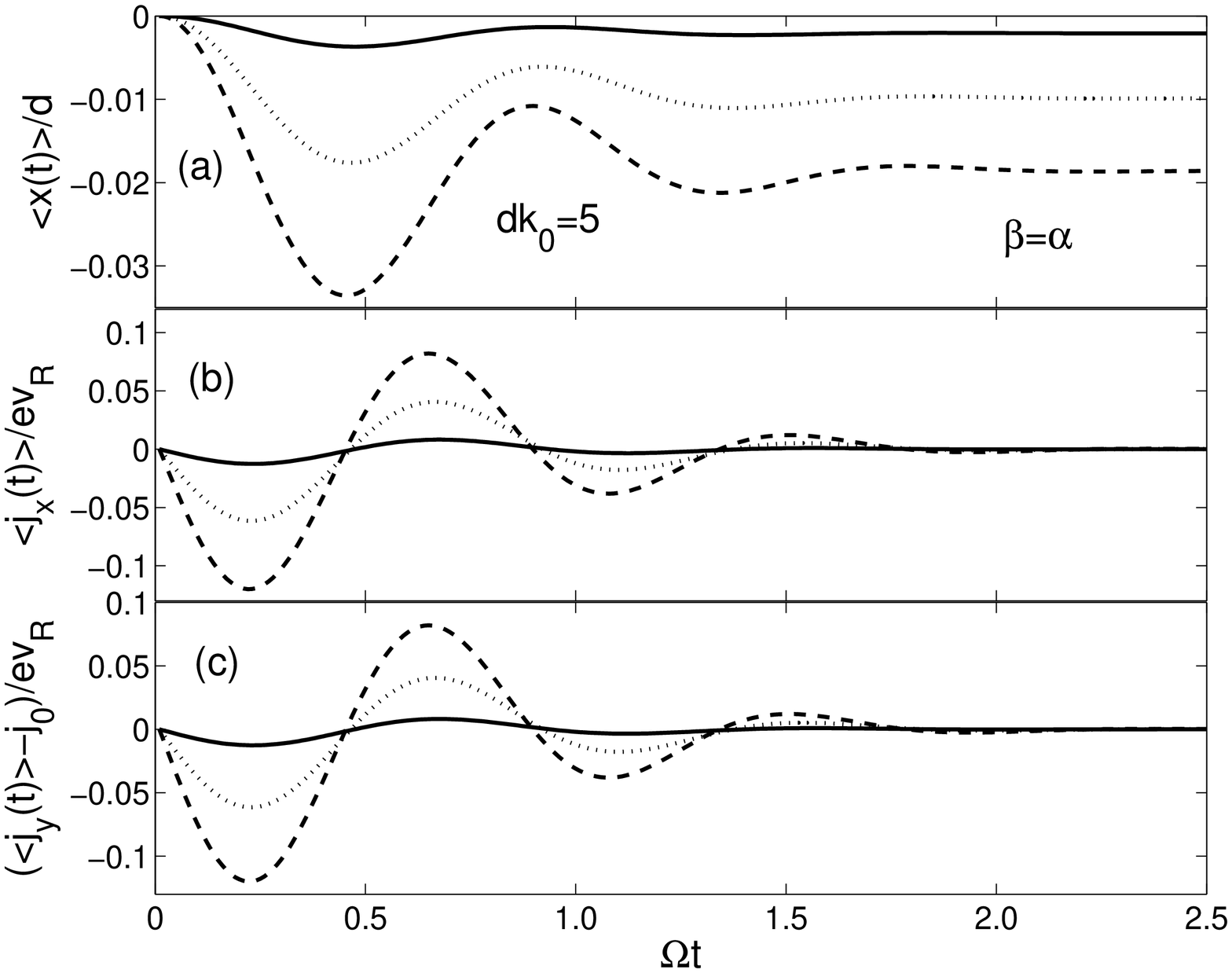}
\caption{Here $\la x(t)\ra/d$, $\la j_x(t)\ra/{ev_{_R}}$ and
$\la j_y(t)-j_0\ra/{ev_{_R}}$ are plotted as a function of
$\Omega t$. In 
all the cases we set $\beta=0.5\alpha$ for GaAs/AlGaAs QW. 
Here, solid line: ($B_x=1/\sqrt{2}$ T, $B_y=1/\sqrt{2}$ T),
dotted line: ($B_x= 3$ T, $B_y=4$ T),
dashed line: ($B_x=6$ T, $B_y=8$ T).}
\label{Fig3}
\end{center}
\end{figure}

\begin{figure}[t]
\begin{center}\leavevmode
\includegraphics[width=105mm]{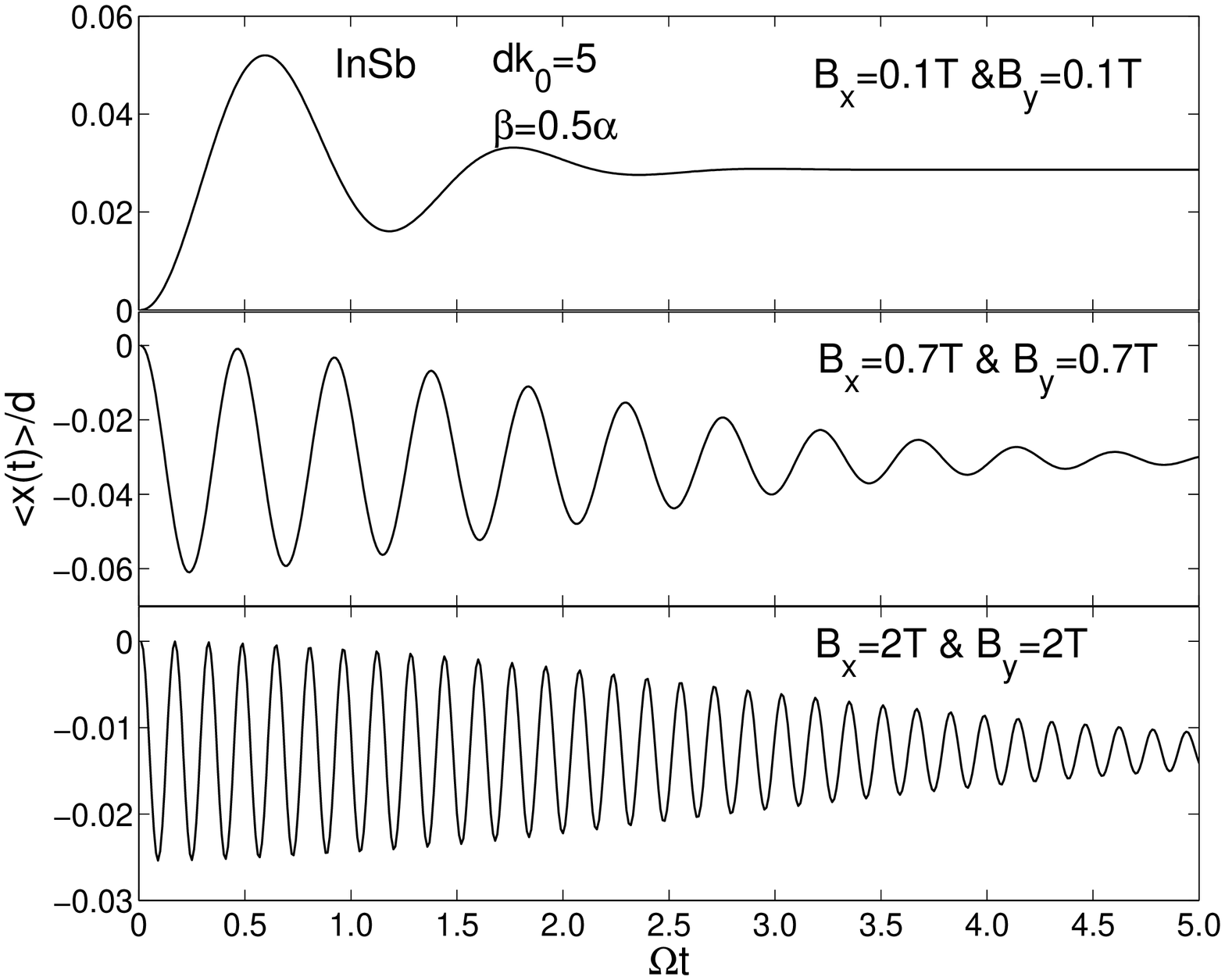}
\caption{ $\la x(t) \ra/d$ is plotted as a function of $\Omega t$ for
InSb QW. Here, we set $\beta=0.5\alpha$.}
\label{Fig4}
\end{center}
\end{figure}

{\bf InSb QW}:
Here, we consider InSb QW for which the 
the effective Lande $g$-factor is very high (e.g $g^\ast=-50$) as 
compared to GaAs/AlGaAs. The RSOI strength is taken to be 
$\alpha=0.9 \times 10^{-11}$ eV-m. We set here $\beta=0.5\alpha$ and $dk_0=5$.
In Fig. [4] $\la x(t) \ra/d$ is plotted with respect to $\Omega t$ for
different values of the magnetic field.
But the situation is different from the GaAs/AlGaAs QW case.
Although the amplitude decreases but the number of oscillations contained
in ZB within the same time range is quite large as we increase the magnetic field. 
Since the magnitude of $g^\ast$ is large the coupling between electron's
spin and magnetic field is strong.

One can obtain more oscillation in ZB 
by increasing the magnitude of the parameter $dk_0$ and we
have shown that when $dk_0>>1$ the pattern is completely 
oscillatory as evident from Eq.(10).
In all these cases it is observed that
the ZB is transient in nature i.e it's amplitude decreases with
time and it is a direct consequence of Lock's \cite{lock}
prediction that the ZB of electron will not be persistent in time
if it is represented by a wave packet.

\section{Zitterbewegung in a Quantum dot}

In this section we would like to study ZB of electrons in a 
semiconductor QD.
We consider a 2DEG confined by an isotropic harmonic 
oscillator potential
$V(x,y)= (1/2)m\omega^2 \big(x^2+y^2 \big)$.
In this context our Hamiltonian reads as
\begin{eqnarray}
H&=&\frac{p^2}{2m}+V(x,y)+
\frac{\alpha}{\hbar}\big(\sigma_x p_y-\sigma_y p_x\big) \nonumber \\
& + & \frac{\beta}{\hbar} \big(\sigma_x p_x - \sigma_y p_y \big) + 
\frac{g^\ast}{2}{\mu_{_B}} \big(B_x\sigma_x + B_y\sigma_y\big).
\end{eqnarray}

We introduce the conventional harmonic oscillator creation 
and annhilation operators as
$a_x = (x/l + i l p_x/\hbar)/\sqrt{2}$,
$a_x^\dagger = (x/l - i l p_x/\hbar)/\sqrt{2}$,
$a_y = (y/l + i l p_y/\hbar)/\sqrt{2}$, and
$a_y^\dagger = (y/l - i l p_y/\hbar)/\sqrt{2}$, where
$l=\sqrt{\hbar/(m \omega)}$ is the harmonic oscillator length.

This Hamiltonian can be re-written as 
\begin{eqnarray}
H &=& \hbar \omega \Big(a_x^\dagger a_x+a_y^\dagger a_y + 1 \Big) \nonumber \\
& + &  \frac{i \alpha}{\sqrt{2} l}
\Big\{ \big(a_x-a_x^\dagger \big)\sigma_y
-\big(a_y-a_y^\dagger\big)\sigma_x\Big\} \nonumber \\
& + & \frac{i \beta}{\sqrt{2} l}
\Big\{ \big(a_y-a_y^\dagger\big)\sigma_y
-\big(a_x - a_x^\dagger\big)\sigma_x\Big\} \nonumber \\
& + & \frac{g^\ast}{2}\mu_{_B} \big(B_x\sigma_x + B_y \sigma_y \big).
\end{eqnarray}

We consider only two lowest occupied energy states 
(ground state and first excited state) of a two-dimensional 
harmonic oscillator potential.
This approximation is known as the ``two sub-band model''. 
Within this approximation the Hilbert space spanned by the following six
basis vectors:
$\vert 0, 0,\uparrow \ra$, $\vert 0, 0, \downarrow \ra$, $\vert 1, 0, \uparrow \ra$,
$\vert 1, 0, \downarrow \ra$, $\vert 0, 1, \uparrow \ra$, $\vert 0, 1, \downarrow \ra$.
Here, $ \uparrow $ and $\downarrow$ represent 
the $z$-component of the electron's spin vector.

Within six basis vectors one can write the Hamiltonian in a matrix form as
\[H=
\begin{pmatrix}
  H_{11} & H_{12} & 0 & H_{14} & 0 & H_{16} \\
  H_{21} & H_{22} & H_{23} & 0 & H_{25} & 0 \\
  0 & H_{32} & H_{33} & H_{34} & 0 & 0 \\
  H_{41} & 0 & H_{43} & H_{44} & 0 & 0 \\
  0 & H_{52} & 0 & 0 & H_{55} & H_{56} \\
  H_{61} & 0 & 0 & 0 & H_{65} & H_{66}
\end{pmatrix}\]
The matrix elements are as follows:
$H_{11}= H_{22} = \hbar \omega = \varepsilon_0$,
$H_{33}= H_{44}=H_{55}=H_{66} = 2\hbar\omega = \varepsilon_1 $,
$H_{12}= H_{21}^{\ast} = H_{34}=H_{56} =  H_{43}^{\ast} = H_{65}^{\ast}
= g^\ast \mu_{_B} \big(B_x-iB_y \big)/2$,
$H_{14} = H_{41}^* = - H_{23}^* = - H_{32} = (\alpha - i \beta)/(l\sqrt{2}) $,
$H_{25}= H_{52}^* = - H_{61} = H_{25}^{\ast} = - H_{16}^* = 
-(\beta + i \alpha)/(l \sqrt{2}) $.

Here $\varepsilon_0=\hbar\omega$ is the zero-point energy 
and $\varepsilon_1=2\hbar\omega$ is the 1st excited state energy of
the two-dimensional harmonic oscillator.

We want to determine the expectation value of the time-dependent 
position operator in this system.
Let us consider at $t=0$ the system is in the ground state and the spin is 
oriented along the positive $z$-direction. This initial state is given by 
$ \vert\psi(0) \ra=\vert0,0,\uparrow\ra $,
so the expectation value of the position operator is given by
\begin{eqnarray}
\la x_{_H}(t)\ra & = & \la \psi(0)\vert x_{_H}(t)\vert\psi(0)\ra\nonumber\\
&=&\la\psi(0)\vert e^{i Ht/\hbar} x(0)
e^{-i Ht/\hbar}\vert\psi(0)\ra\nonumber\\
&=&\la0\vert VV^\dagger e^{i Ht/\hbar}VV^\dagger xVV^\dagger 
e^{-i Ht/\hbar}VV^\dagger\vert0\ra\nonumber\\
&=&\la0\vert VV^\dagger e^{i Ht/\hbar}V X V^\dagger 
e^{-iHt/\hbar}VV^\dagger\vert0\ra\nonumber\\
&=&\sum_{i=1}^{6}\Big(\sum_{j=1}^{6}V_{j1}H_{ij}
e^{i\frac{(\lambda_i-\lambda_j)t}{\hbar}}\Big)V_{i1}^\dagger,
\end{eqnarray}
where $V$ is the diagonalization matrix which diagonalizes 
the Hamiltonian $H$ and $VV^\dagger=I$. Also,
$ \omega_{ij} = (\lambda_i-\lambda_j)/\hbar $ is the beating
frequency.  

The $V$ matrix also diagonalizes $U(t)$ and becomes
$U_{diag}(t)=VU(t)V^\dagger=e^{-i \lambda_i t/\hbar}I_{ij}$ and
$X=VxV^\dagger$.
The position operator is given by
$x = \lambda \big(a+a^\dagger \big)/\sqrt{2} $.
Within the above mentioned basis this can be written in a matrix form as

\[x=\frac{\lambda}{\sqrt{2}}
\begin{pmatrix}
  0 & 0 & 0 & 0 & 1 & 0 \\
  0 & 0 & 0 & 0 & 0 & 1 \\
  0 & 0 & 0 & 0 & 0 & 0 \\
  0 & 0 & 0 & 0 & 0 & 0 \\
  1 & 0 & 0 & 0 & 0 & 0 \\
  0 & 1 & 0 & 0 & 0 & 0
\end{pmatrix}\]

\begin{figure}[ht]
\begin{center}\leavevmode
\includegraphics[width=105mm]{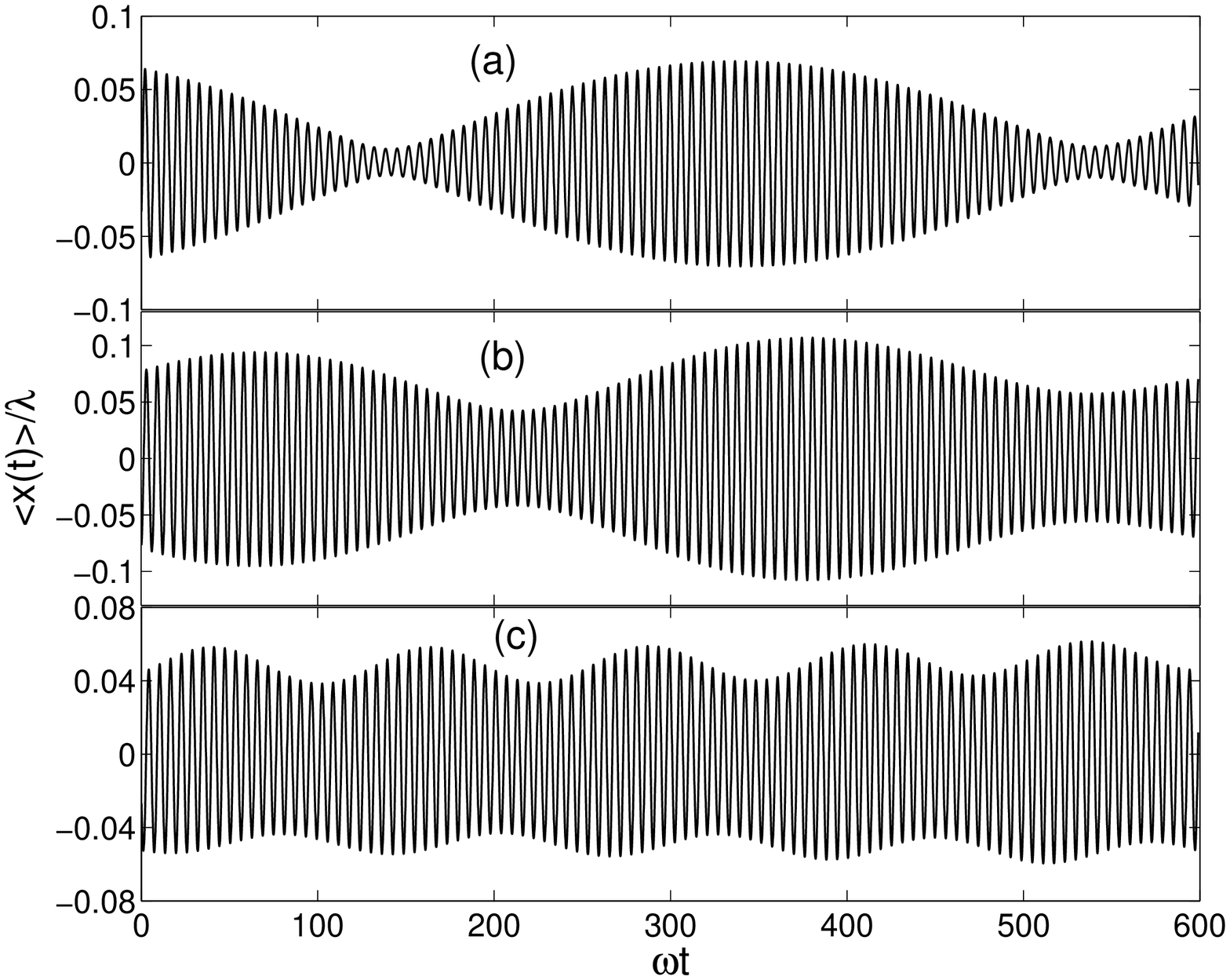}
\caption{Plots of $\la x(t)\ra/\lambda$ 
vs $\omega t$ for GaAs/AlGaAs QD for different values of 
magnetic fields. Here, Fig. (a): ($B_x=0.1$ T, $B_y=0.1$ T),
Fig. (b): ($B_x=1/\sqrt(2)$ T, $B_y=1/\sqrt(2)$ T),
Fig. (c): ($B_x=6$ T, $B_y=8$ T). We consider only RSOI i.e. $\beta=0$.}
\label{Fig8}
\end{center}
\end{figure}

\subsection{Numerical Results and Discussion}
The ZB of a GaAs/AlGaAs quantum dot in the presence of the in-plane
magnetic field with SOI is investigated here. 
The value of the Rashba strength is taken as
$\alpha = 1.0 \times 10^{-11} $ eV-m and 
the zero-point energy of the harmonic oscillator potential is fixed to 
the value $\varepsilon_0=5$ meV. We find the expectation values of 
the position coordinate as a function of time which are plotted in 
Figs. [5] and [6]. In Fig. [5] we fix $\beta=0$ and vary the magnetic 
field strengths.
The magnetic field is kept constant and $\beta$ is varied in Fig. [6].
The ZB in quantum dots is similar to the beating effect in the classical
wave mechanics with different frequencies. The oscillatory motion is due to 
the superposition of individual oscillatory motions with frequencies 
corresponding to the all possible energy eigenvalue differences of the 
Hamiltonian.
In this case, there are six non-degenerate eigenvalues and we have six
values of the energy differences. The number of beating frequencies is 
six.

\begin{figure}[t]
\begin{center}\leavevmode
\includegraphics[width=105mm]{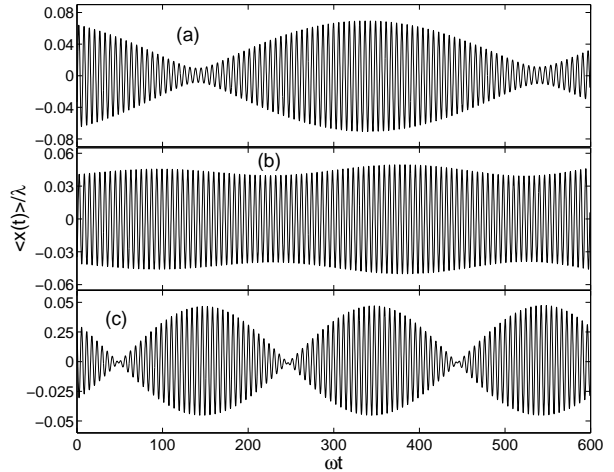}
\caption{Plots of $\la x(t)\ra/\lambda$ 
vs $\omega t$ for GaAs/AlGaAs QD for different values of $\beta$.
The value of magnetic field is kept fixed to $(B_x,B_y)=(0.1,0.1)T$.
Figures (a),(b) and (c) correspond to $\beta=0$, $\beta=0.5\alpha$
and $\beta=\alpha$ respectively.}
\label{Fig9}
\end{center}
\end{figure}

\section{Conclusion}
In this work we have investigated the effect of an 
external in-plane magnetic field on the ZB of an electron 
in semiconductor QW and QD with Rashba and 
Dresselhaus spin-orbit interactions. For QW a general expression of 
the expectation values 
of position coordinate and current due to ZB within the 
Gaussian wave packet is obtained. 
For QW case, the oscillatory quantum motion of electron 
which is represented by a wave packet shows transient 
behavior and this signature is a proof of Lock's argument. 
Another important point is that
ZB does not vanish even at $\alpha=\beta$ when a finite in-plane magnetic 
field is present. The $y$-component of current also performs ZB motion 
with finite magnetic field.
We study the same problem for high Lande g-factor QW like InSb 
in comparison with low Lande g-factor QW like GaAs-AlGaAs.
We have also studied the problem of ZB in a GaAs/AlGaAs QD numerically.
The ZB in GaAs/AlGaAs QD is persistent in time. 
The ZB in quantum dots shows beating-like pattern and it is similar to the
the beating effect in the classical wave mechanics.

\end{document}